\documentclass[twocolumn,showpacs,aps,pra]{revtex4}
\usepackage{graphicx,amsmath,amssymb,natbib,epsfig}

\begin{document}
\title{Method for traveling-wave deceleration of buffer-gas beams of CH}
\author{M.I. Fabrikant$^1$, Tian Li$^2$\footnote{Present Address: Joint Quantum Institute, National Institute of Standards and Technology and the University of Maryland, College Park, MD 20742, USA}, N.J. Fitch$^1$, N. Farrow$^1$, Jonathan D. Weinstein$^2$, and H.J. Lewandowski$^1$}

\affiliation{$^1$JILA and Department of Physics, University of Colorado, Boulder, Colorado 80309-0440\\$^2$Department of Physics, University of Nevada, Reno, Nevada 89557, USA}

\date{\today}

\begin{abstract}
Cryogenic buffer-gas beams are a promising method for  producing  bright sources of cold molecular radicals for cold collision and chemical reaction experiments.
 In order to use these beams in studies of reactions with controlled collision energies, or in trapping experiments, one needs a method of controlling the forward velocity of the beam. A Stark decelerator can be an effective tool for controlling the mean speed of molecules produced by supersonic jets, but efficient deceleration of buffer-gas beams presents new challenges due to longer pulse lengths. Traveling-wave decelerators are uniquely suited to meet these challenges because of their ability to confine molecules in three dimensions during deceleration and their versatility afforded by the analog control of the electrodes. We have created ground state CH$(X^2\Pi)$ radicals in a cryogenic buffer-gas cell with the potential to produce a cold molecular beam of $10^{11}$ mol./pulse.  We present a general protocol for Stark deceleration of beams with a large position and velocity spread for use with a traveling-wave decelerator. Our method involves confining molecules transversely with a hexapole for an optimized distance before deceleration. This rotates the phase-space distribution of the molecular packet so that the packet is matched to the time varying phase-space acceptance of the decelerator. We demonstrate with simulations that this method can decelerate a significant fraction of the molecules in successive wells of a traveling-wave decelerator to produce energy-tuned beams for cold and controlled molecule experiments.
\end{abstract}


\maketitle

\section{Introduction}

Cryogenic buffer-gas methods have the potential to create bright beams of a large range of cold molecule species that have not yet been studied in depth because of the challenge of creating supersonic jets of these molecules\cite{Doyle2012BeamReview, Heller2012}. There are several applications of intense molecular beams, from experiments attempting to measure the electron's electric dipole moment \cite{acme,fountain} and variation of fundamental constants \cite{ImperialCH}, to experiments exploring cold reactions relevant to interstellar cloud chemistry \cite{insterstellar}. In most of these experiments, removing the mean forward velocity is critical to taking full advantage of the molecular source. Over the last ten years, many methods for molecular deceleration have been developed, with Stark deceleration being the most widely used technique \cite{Bigreview}. However, no one has yet used a Stark decelerator to decelerate these intense buffer-gas beams. Thus far, experimental approaches for decelerating such beams has been limited to direct laser slowing \cite{SRF} and combining magnetic potentials with optical pumping \cite{Doylemagload2013}.

To demonstrate the advantages of a combined system of a buffer-gas beam with a Stark decelerator, we chose CH (methylidyne), the simplest organic molecule.  CH is difficult to create in the laboratory because it is a tri-radical.  However, it plays an important role in interstellar medium chemistry and combustion chemistry. CH participates in archetypical reactions such as hydrogen exchange with deuterium \cite{soft2009a}
\begin{equation}
\mathrm{CH}(X^2\Pi)+\mathrm{D}_2\rightarrow \mathrm{CD}(X^2\Pi)+\mathrm{HD},
\end{equation}
the formation of more complex hydrocarbons \cite{Brownsword97}
\begin{equation}
\mathrm{CH}+\mathrm{H}_2\rightarrow \mathrm{CH}_3+h\nu,
\end{equation}
and simple combustion reactions \cite{C0CP01529F}
\begin{align}
	\begin{split}
	\mathrm{CH}(X^2\Pi)+\mathrm{C_2H_2}\rightarrow \mathrm{C_3H_2} +\mathrm{H} \\
        \mathrm{CH}(X^2\Pi)+\mathrm{C_2H_2}\rightarrow \mathrm{C_3H+H_2}.
	\end{split}
\end{align}

In the cold, dilute, interstellar medium, the most important reactions can be expected to be two-body barrierless reactions \cite{smithreview}. In order to study such reactions, the ability to prepare the reactants with extremely well known interaction energies is crucial. CH reactions have been studied in crossed beam experiments \cite{Zhang2011}, and at temperatures as low as 23~K \cite{Loison2009}, but collision experiments in which the CH beam is both internally cold and traveling at a low velocity remain unexplored. We propose that progress toward such experiments can be made by combining a molecular beam of CH with a Stark decelerator, which will allow us to create a cold, bright, controlled velocity source of CH. In addition, molecules in a decelerated CH beam could potentially be combined and trapped with magnetic or electric fields. This would enable the study of collisions down to energies in the 10-100 mK regime with essentially only one quantum state populated \cite{Fitch20121}.

This paper describes experimental work investigating the production of CH in a cryogenic buffer-gas cell, and detailed calculations of the coupling of a cryogenic beam of CH to a traveling-wave Stark decelerator.

\section{Cryogenic production of CH}
We produce methylidyne (CH) by laser ablation of a solid target and cool it with a cryogenic helium buffer gas.
The experimental apparatus and techniques are as described in Ref.\cite{UNR08TiHe}.  The cryogenic cell in which the experiment takes place is modified from that reference to have an internal volume of roughly 10~cm~$\times ~10$~cm~$\times ~ 2.5$~cm.

We chose iodoform (CHI$_3$) as the solid precursor in the hopes that the weak C---I bond would favor the formation of CH in ablation, inspired by prior work producing CH from photolysis of gas-phase bromoform \cite{Butler1979104}.  We obtained iodoform  in powder form; to form suitable targets for ablation, we dissolved iodoform powder (99\% purity) in acetone and let the acetone evaporate to leave a solid  on a metal substrate. However, the targets that produced the observed signal had turned a blackish color after evaporation (changed from the original yellowish powder color), suggesting that the chemical composition of the ablation target is no longer pure iodoform.

\begin{figure}[ht]
    \begin{center}
    \includegraphics[width=\linewidth]{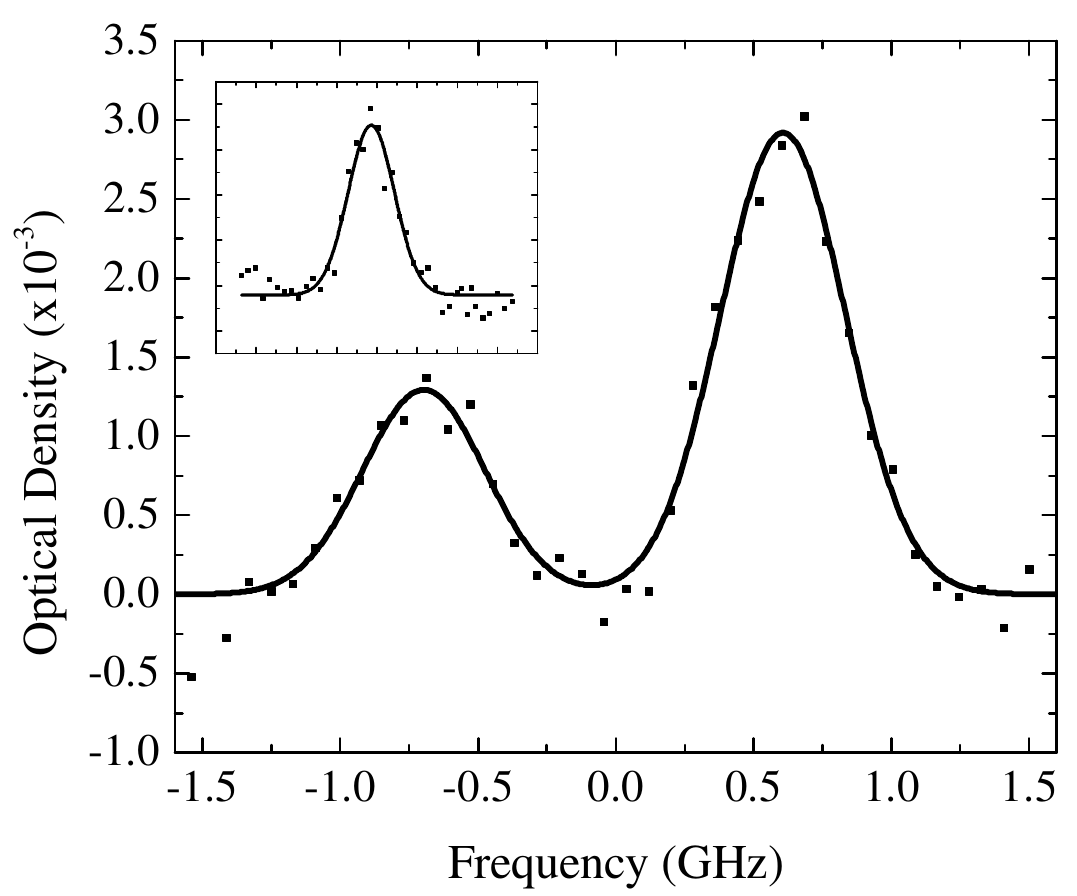}
    \caption{ \label{fig:CH_spectra}
 Spectra of $ X^2\Pi (v''=0, N''=1, J''=1/2)$  cryogenic CH molecules.  The main figure shows the $Q$-branch transitions;
  the frequency offset is 25723.4~cm$^{-1}$.
  The inset shows the $P$-branch transition plotted on the same scale; its frequency offset is 25698.2~cm$^{-1}$.
  The spectra were obtained from 1 to 2~ms after the ablation pulse.  The experimental measurements are shown as points, the fit to a Gaussian (two Gaussians, in the case of the $Q$-branch transition) is shown as a solid line.  The data were taken at a cell temperature of 5~K, ablation energy of 0.1~J, and helium buffer gas density of $1 \times 10^{16}$~cm$^{-3}$.
   %
%
    }
    \end{center}
\end{figure}

We detect the cold methylidyne molecules by laser absorption spectroscopy.  Typical probe beam powers are on the order of a few microwatts, with a probe beam diameter of a few~mm.
We observe CH molecules in the $X ^2\Pi (v''=0, N''=1, J''=1/2)$ rovibrational ground state on the $B ^2 \Sigma^- \leftarrow X^2\Pi$ $Q$- and $P$-branch transitions at 389~nm
\cite{Kepa1996}.
The spectra are shown in Fig. \ref{fig:CH_spectra}.
We note that we are unable to resolve the ground-state hyperfine structure and that parity selection rules prevent measurement of the lambda-doubling in the ground state \cite{McCarthy2006}.

The two peaks observed on the $Q$-branch transition 
are due to the spin-rotation splitting of the $J'=1/2$ and $J'=3/2$
 states of the $B ^2\Sigma (N'=1)$ excited state.
The $P$-branch transition to the $B ^2\Sigma (N'=0, J'=1/2)$  state shows only a single absorption peak.
 The relative heights of the three peaks are consistent with calculated absorption coefficients \cite{luque1996}.

 From these spectra, we calculate that we produce $2\times 10^{11}$ CH molecules in each of the lambda-doublet states of   $X ^2\Pi (v''=0, N''=1,  J''=1/2)$ \cite{luque1996}.
Similar numbers are observed for helium buffer gas densities from $4 \times 10^{15}$ to $1 \times 10^{17}$~cm$^{-3}$.  We note that high-flux cryogenic buffer-gas beam sources typically employ buffer-gas densities from $10^{15}$ to $10^{16}$~cm$^{-3}$ \cite{DoylePCCP2011}.

 The measured temperature for the data shown in Fig. \ref{fig:CH_spectra} is  $17 \pm 2$~K, significantly higher than the 5 K cell wall temperature measured prior to ablation.  This can be attributed to the large ablation power and short observation time after the laser pulse \cite{PhysRevA.83.023418}.  Lower ablation powers of 0.05~J showed temperatures of $13 \pm 1$~K.  While this increase in temperature might be quite deleterious for a helium-based beam source intended to operate at a temperature of a few kelvin, it will be of little adverse consequence for a neon-based beam source designed to operate at temperatures approaching 20~K \cite{DoylePCCP2011}.

\begin{figure}[ht]
    \begin{center}
    \includegraphics[width=\linewidth]{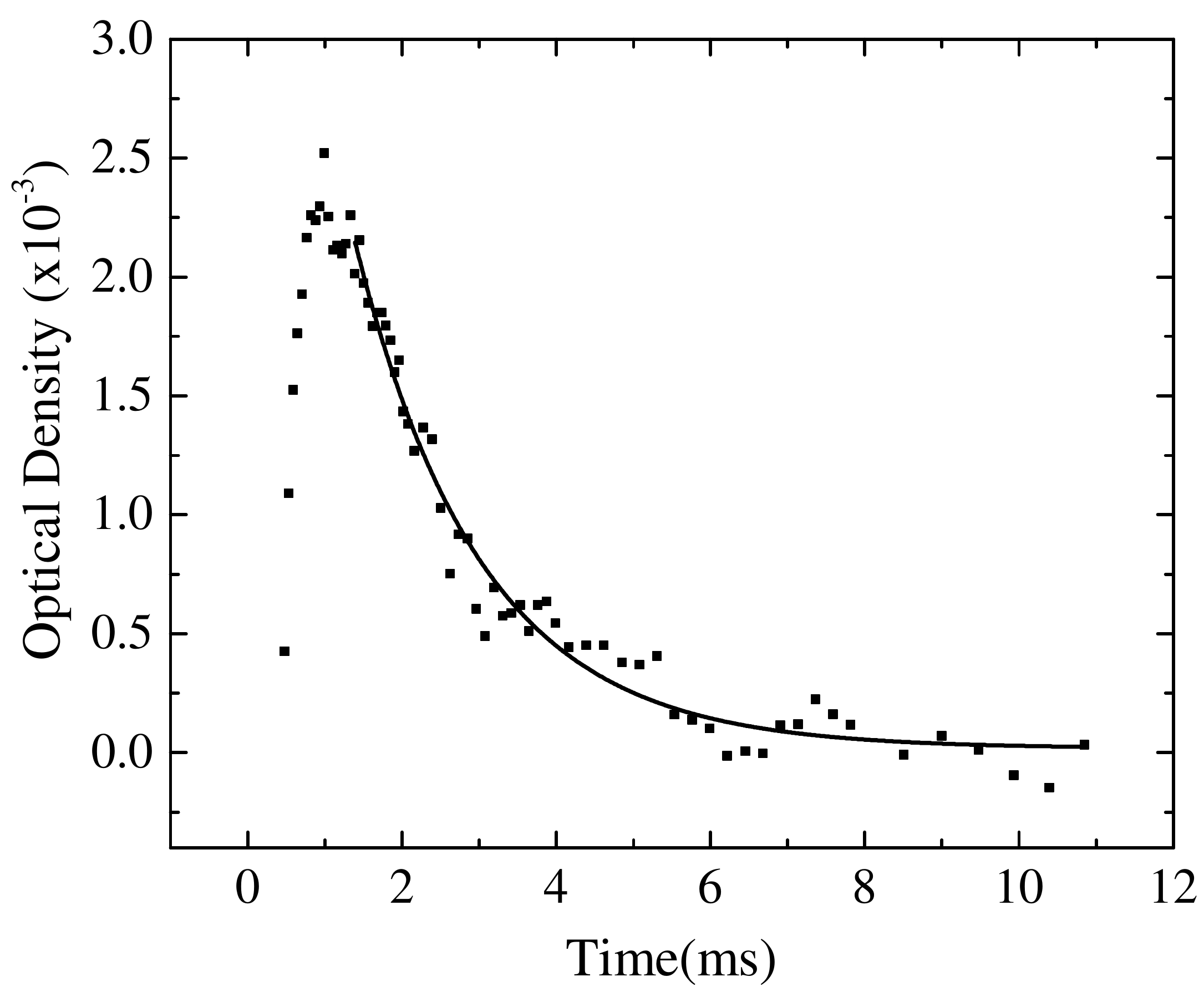}
    \caption{ \label{fig:CH_OD_vs_time}
    Measured optical density on the Q-branch transition as a function of time after the laser ablation pulse.
    The experimental measurements are shown as solid squares, the fit to an exponential  is shown as a solid line; the exponential time constant is 2~ms.  The data were taken at a cell temperature of 5~K, ablation energy of 0.1~J, and helium buffer gas density of $1 \times 10^{16}$~cm$^{-3}$.
    }
    \end{center}
\end{figure}

The temporal behavior of the CH signal is shown in Figure \ref{fig:CH_OD_vs_time}.  At long times after the ablation pulse, the optical density is observed to decrease exponentially in time, as expected for diffusion.  In addition, the exponential lifetime increases linearly with helium density over the range from $4 \times 10^{15}$ to $5 \times 10^{16}$~cm$^{-3}$, indicating that the dominant loss mechanism is diffusion to the cell walls, and not chemical reactions with other species produced by the ablation.

\section{Stark Deceleration of a position/velocity correlated beam}
Stark deceleration is a method that uses time varying inhomogeneous electric fields to reduce the mean velocity of a molecular beam via the interaction of an electric field with a molecule's dipole moment. Previously, this had been realized as a spatially periodic array of high-voltage electrode pairs that create electric field maxima along the molecular beam path \cite{PhysRevLett.106.193201}. These maxima create a time-averaged potential well that periodically removes kinetic energy from the target molecules. This method has been described previously in Ref. \cite{Meijer2012}. Recently, a new method of Stark deceleration has been demonstrated in which a series of ring electrodes, with continuously varying voltages applied, create a true moving potential well that decelerates molecules \cite{OrignalRing}. The potential well moves initially at the mean speed of the molecular beam, but subsequently decreases its speed to decelerate molecules in the trapped potential well. This method has been implemented both alone \cite{Perez2013} and in combination with a traditional decelerator \cite{Bethlem2012}. The salient advantage of this ``traveling wave" decelerator  is the true three dimensional confinement of the molecules, which inhibits transverse losses during the deceleration process.

Stark deceleration of supersonic beams has been studied in great detail \cite{Meijer2012}, but its application to the different beam parameters offered by a cryogenic buffer-gas source remains unexplored.  Typical parameters for the two sources have been well documented \cite{doyle05beam, Doyle2012BeamReview, 1306.0241}. For our simulations, we chose parameters listed in Table. \ref{thetable}, typical for a neon buffer-gas beam in the hydrodynamic expansion regime. \cite{DoylePCCP2011}.

\begin{table}\label{thetable}
\begin{tabular}{c c c}
\hline

Average Forward Velocity&$180$~m/s\\
Longitudinal Velocity Spread&17~m/s ($1\sigma$)\\
Longitudinal Position Spread&10 cm ($1\sigma$)\\
Transverse Velocity Spread&21 m/s ($1\sigma$)\\
Transverse Position Spread&1.7 mm ($1\sigma$)\\
\hline
\end{tabular}
\caption{Parameters of the buffer-gas beam used in simulations. The expansion out of the cell is assumed to be in the hydrodynamic regime. These parameters would correspond to a cell extraction time of $\sim$ 2 ms and a cell aperture diameter $\sim$ 5 mm for a physical cell \cite{DoylePCCP2011}.
}
\label{stark}
\end{table}

A buffer-gas source can create beams with transverse and longitudinal velocity spreads similar to those of a supersonic beam, but with reduced longitudinal velocities \cite{DoylePCCP2011}.  The lower mean speed makes deceleration less technically demanding as the frequency of the changing potentials is reduced.  However, the large temporal spread of buffer-gas sources creates large longitudinal position spreads, which means only a small fraction of the beam will fit inside a single potential well of the decelerator. Also, the spread in longitudinal velocity is much larger than the velocity acceptance of the decelerator. To address these two problems, we developed a protocol to correlate the longitudinal position and velocity of the molecular beam and to match the deceleration of the potential wells to that correlated phase-space distribution. This results in molecules being loaded into many successive wells of the decelerator and thus a considerable fraction of the beam being decelerated.

\begin{figure}[ht]
    \begin{center}
    \includegraphics[width=\linewidth]{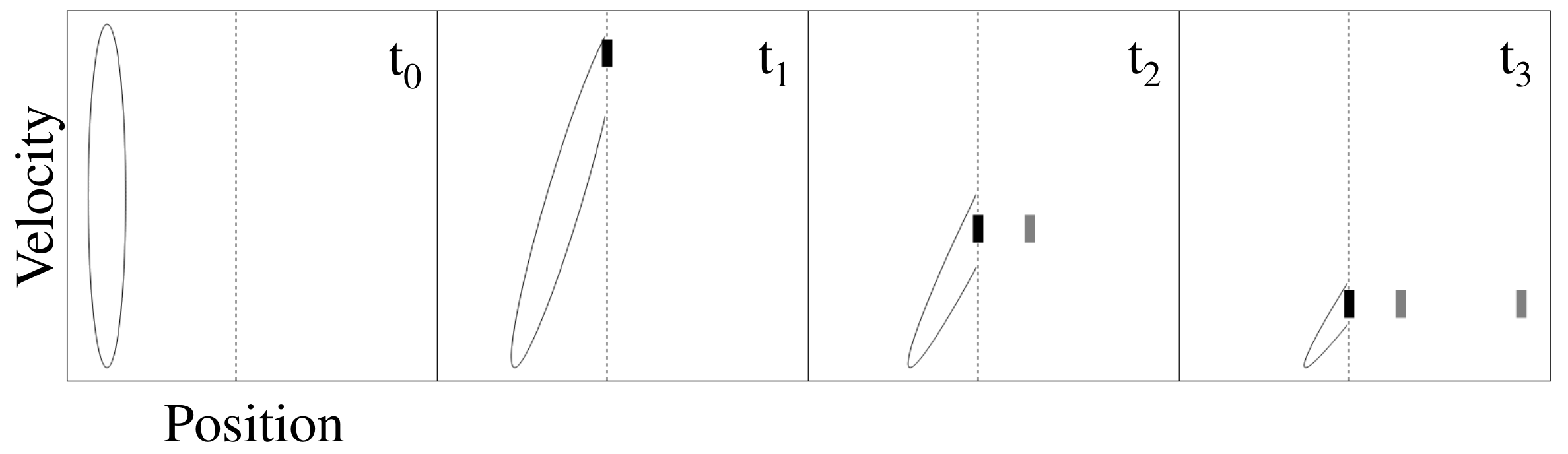}
    \caption{ \label{Accepting}
    Time series schematic of the molecule acceptance scheme. As the molecular packet(ellipse) approaches the decelerator entrance (represented by the dotted line) its longitudinal phase space rotates and stretches. The velocity of molecules arriving at the decelerator entrance decreases over time. If the decelerator well velocity changes in a manner that matches the changing molecule arrival velocity, molecules can be captured and decelerated in successive potential wells (rectangles).}
    \end{center}
\end{figure}

    The correlation of the velocity and position of the molecule beam is accomplished by allowing the beam to propagate through a long hexapole before entering the decelerator. During this propagation, the molecules with higher longitudinal speeds move ahead of the center of the packet, while the slower molecules lag behind, creating a position/velocity correlation\cite{Meek26062009}. The instantaneous potential well velocity is then chosen to match the velocity of arriving molecules, dynamically changing the decelerator's phase-space acceptance (PSA) to match the phase-space distribution (PSD) of the arriving molecular beam. In this way, many wells of the decelerator can be loaded with a high density of molecules. This approach is illustrated schematically in Figure  \ref{Accepting}.

To establish how the velocity of the decelerator wells should change in time, we consider the velocity of a molecule as it enters the decelerator after being guided from the source aperture by a hexapole. In the limit of a long hexapole guide, a molecules's longitudinal speed at the decelerator entrance will vary as $v = H/t$, where $H$ is the length of the hexapole, and $t$ is time the packet has been propagating.
In this limit, one would achieve perfect coupling if the velocity of the decelerator's potential wells, $V_a$, had the same functional form. Using this exact form is impractical, since changing $V_a$ as $1/t$ requires an infinitely long decelerator to decelerate molecules to rest.  A more practical deceleration protocol is a linear chirp function, which approximates the “ideal” acceleration function.

Because of the wide velocity spread in our beam, the optimal choice of acceleration in a linear velocity scheme is not immediately obvious. We define the acceleration through the use of an ``index molecule.'' The index molecule velocity determines the acceleration, $a$ by
\begin{equation}
a=\frac{(V_f^2-V_i^2)}{2S}
\end{equation}
where $V_i$ is the velocity of the index molecule, $V_f$ is the final velocity after deceleration, and $S$ is the length of the decelerator. We then change the velocity of the decelerator wells according to
\begin{equation}
V_a(t)=-a(t-\frac{H}{V_i})+V_i
\end{equation}

We investigate the effect of the different deceleration protocols and parameters both by using 3D Monte Carlo simulations and by developing a simple 1D model.

\section{Molecular Trajectory Simulations}

We begin simulations by calculating the position and time dependent Stark energies of a CH molecule within the decelerator using a commercial finite element solver. Stark energies are calculated for sinusoidal voltages applied to the ring electrodes with a peak-to-peak amplitude of 24 kV. The ring electrodes have an inner diameter of 4 mm, a wire diameter of 1~mm, and are spaced by 2~mm. The 4~mm inner diameter hexapole is modeled with an ideal potential with adjacent rods having a 400 V potential difference.  The hexapole voltage was optimized in simulations for best transverse phase-space matching to the decelerator, and is the same for all guide lengths. The hexapole potential also features a hard cutoff at the rod radius, outside of which molecules exit the simulation. Next, we generate a gaussian-distributed molecule packet in all 6 dimensions of phase space using parameters listed in Table I, which are meant to reflect typical pulse parameters reported in the literature \cite{DoylePCCP2011,DoyleChem2012}. We evolve the trajectories of 40,000 molecules through the decelerator with a standard integrator and record their final position in phase-space.

\begin{figure}[ht]
    \begin{center}
    \includegraphics[width=\linewidth]{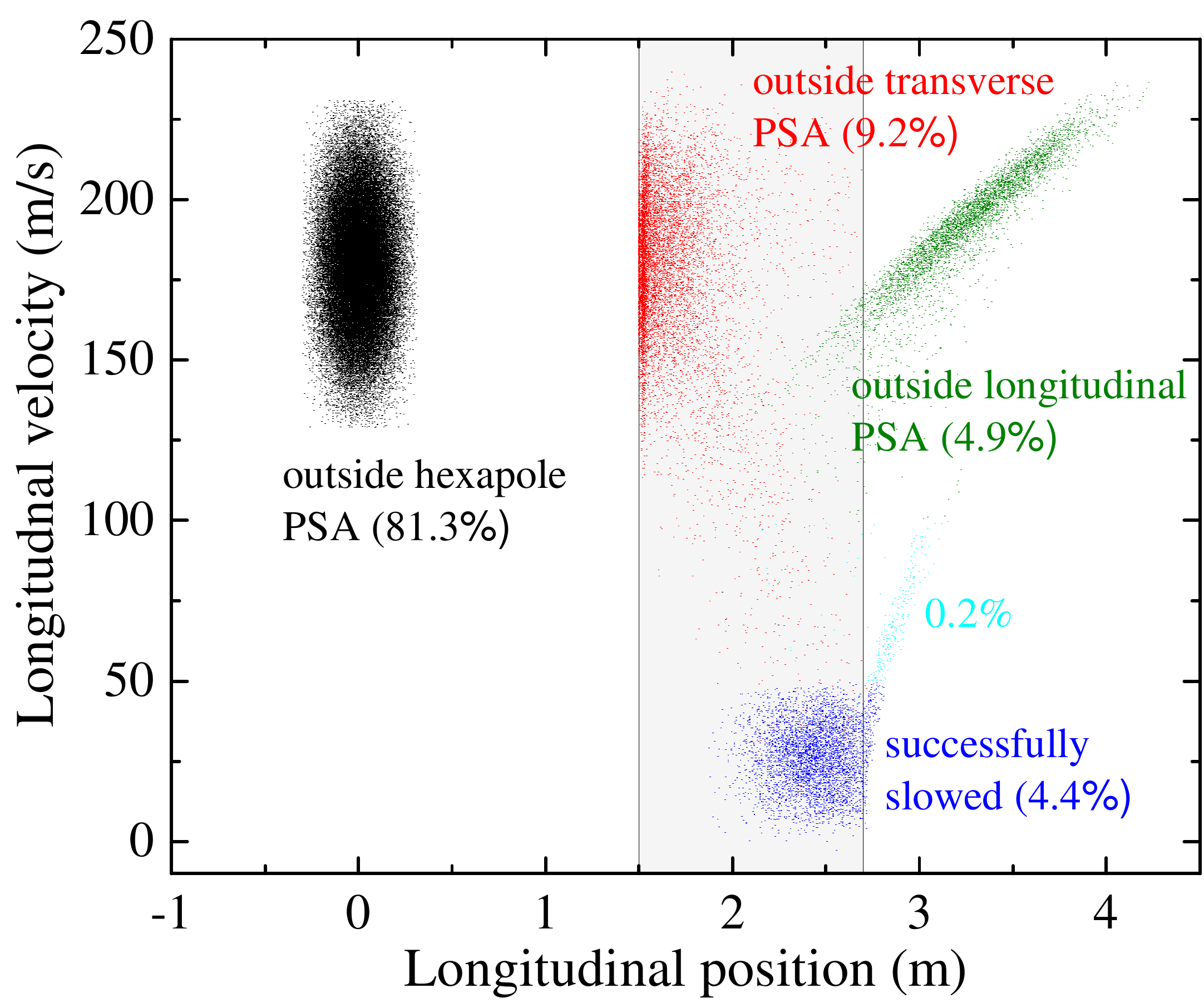}
    \caption{ \label{FinalPhase}
    (Color online) Simulated final longitudinal phase-space distribution for linear deceleration using a 1.5~m hexapole and 210 m/s index molecule. Molecules are frozen at their phase-space coordinates when they exit the simulation by moving beyond the inner diameter of the hexapole or slower or the simulation ends.  The shaded region represents the location of the decelerator.
    }
    \end{center}
\end{figure}

We can understand much of what goes on during deceleration by examination of the final molecular phase-space distribution. In Fig. \ref{FinalPhase}, molecules have been decelerated from 180 m/s to 25 m/s with a pre-decelerator hexapole length of 1.5 m. The final velocity of 25~m/s was chosen because it is a typical value for loading molecules into an external trap. Molecules that leave the simulation by going beyond the hexapole or decelerator ring radius have their phase-space coordinates recorded and cease to evolve, and are tagged with their loss mechanism.

\begin{figure}[ht]
    \begin{center}
    \includegraphics[width=\linewidth]{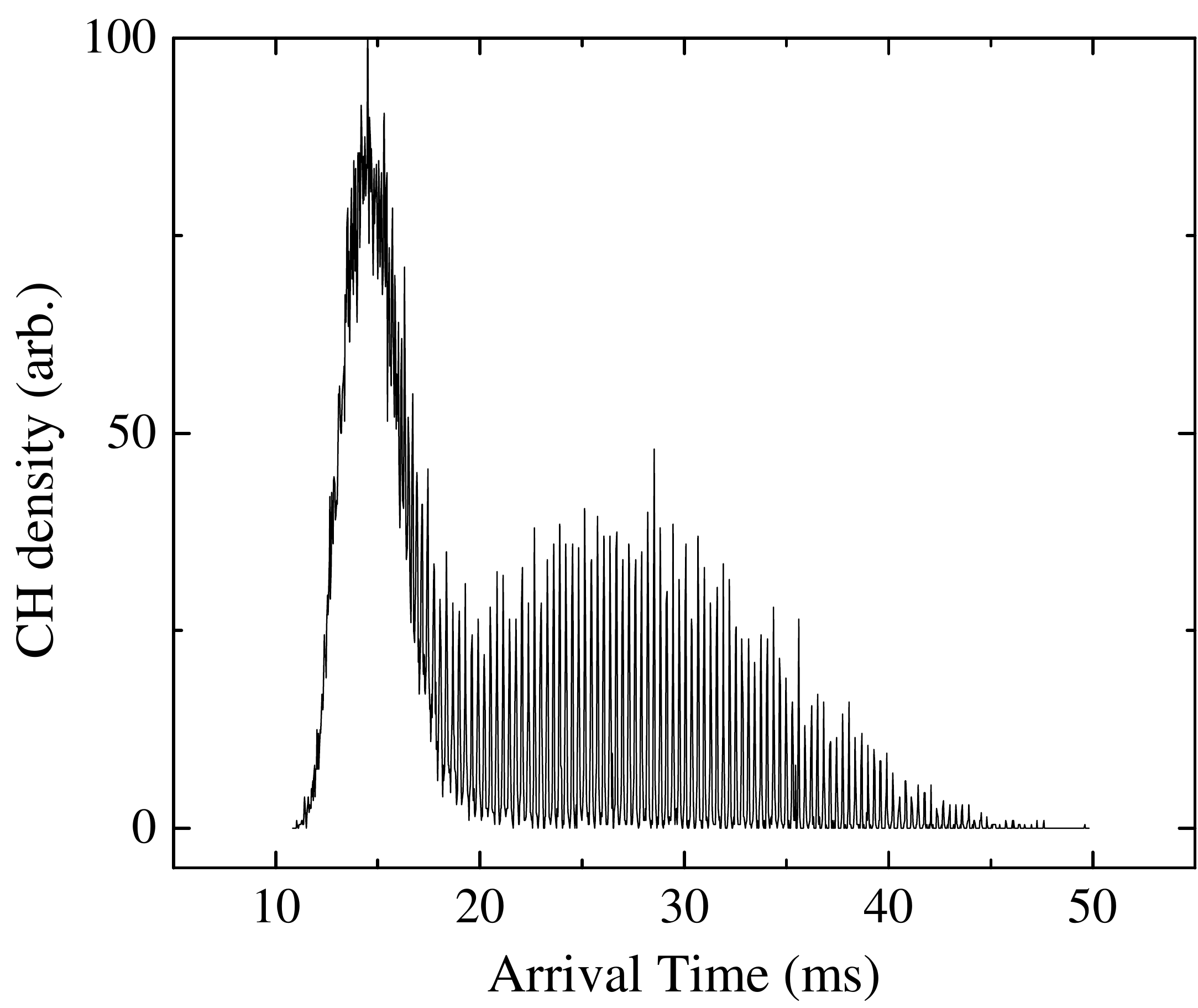}
    \caption{ \label{TOF} Simulated time-of-flight trace for deceleration of a buffer-gas beam. The parameters are the same as those for Fig. \ref{FinalPhase}.
    }
    \end{center}
\end{figure}

Because of the large aperture of the buffer gas cell, most molecules start outside the transverse PSA of the hexapole, and are thrown out immediately (black points).
Molecules that are not accepted into the decelerator either hit electrodes at the beginning of deceleration (red points) or are longitudinally phase unstable and emerge from the decelerator undecelerated (green points). There are also a small number of molecules (0.2\%) that are phase stable, but exit the decelerator before the wells reach the final velocity of 25~m/s. The final phase-space plot also reveals a striking consequence of working with a beam with such a large longitudinal width:  the decelerated molecules are distributed over a length of $\sim$0.5~m. In a time-of-flight trace, the decelerated molecules arrive in dozens of packets (depending on deceleration parameters), as can be seen in Fig. \ref{TOF}.

The full molecular dynamics calculations of molecular trajectories used to obtain the results shown in Fig. \ref{FinalPhase} and Fig. \ref{TOF} have been shown to give excellent agreement with other experiments \cite{Lew2009} and thus are good predictors of possible experiments described here.  However, they are computationally demanding, and thus not ideal for optimizing experimental parameters.  As described below, we developed a 1D model of the beam deceleration.  Although it is unable to yield absolute numbers of decelerated molecules, we have shown that it provides accurate calculations of \emph{relative} numbers for different decelerator parameters, and is much less computationally demanding than the full simulations.  This is possible for a traveling-wave decelerator because of the near-perfect decoupling of longitudinal and transverse motion in the decelerating potential wells.

\subsection{A 1D Model of Phase-Space Acceptance Matching to a Molecular Beam}
The one-dimensional model is based on the overlap between the molecular packet's PSD and the time varying PSA of the decelerator. We neglect the transverse phase-space dimensions because the transverse acceptance does not vary significantly between the different deceleration protocols we evaluated. The PSD of the molecular packet is modeled as a bivariate Gaussian distribution $G(z, V_z, t)$, where $z$ and $V_z$ are the position and velocity variables. At the time the decelerator turns on, $t_0$, the number of molecules within the PSA of the decelerator is approximated by

\begin{equation}
n=\int_{V_a-\Delta v/2}^{V_a+\Delta v/2} G(H,V_z,t_0)\text{ }dV_z(t_0),
\end{equation}
where $\Delta v$ is the width of the decelerator acceptance in the velocity coordinate. $\Delta v$ is a function of the acceleration used, which is the time derivative of $V_a$.

Because molecules continue to be loaded into the decelerator after $t_0$, we must integrate over all later times, resulting in the double integral for the total number of molecules decelerated given by

\begin{equation}\label{ModelIntegral}
N=\int_{t_0}^{t=\infty}\int_{V_a(t)-\Delta v/2}^{V_a(t)+\Delta v/2} G(H, V_z, t)\text{ }dV_z(t) \mathrm{d}t.
\end{equation}

 We note that this model neglects losses within the decelerator.  This omission is justified by the 3D simulations, as seen in Fig. \ref{FinalPhase}. Also, in actual deceleration experiments and in simulations, molecules are accepted in small bunches, but the 1D model incorporates this changing acceptance as a continuous function.

 1D models generally fail to accurately predict the results of the deceleration process of traditional pulsed decelerators because of coupling between the longitudinal and transverse motions \cite{PhysRevA.73.023401}. 3D simulations have been shown to accurately reproduce the molecular trajectories in detail \cite{Lew2009}. We show here that for traveling-wave decelerators a much less computationally intensive 1D model reproduces the results of the full 3D simulations.

\subsection{Results}

\begin{figure*}[ht]
    \includegraphics[width=\linewidth]{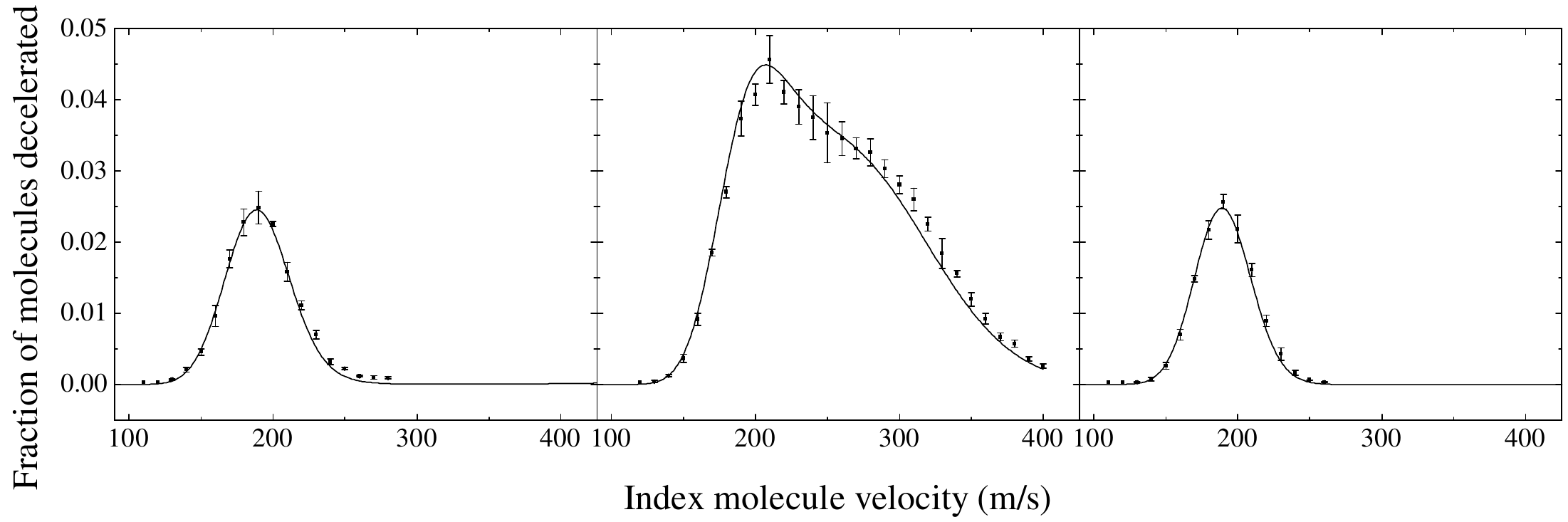}
    \caption{ \label{AllHex}
    The fraction of molecules decelerated using a linear chirp for hexapole guide lengths of 0.5~m (left), 1.5~m (center), and 4~m (right). The results of 3D simulations are shown as black squares and the solid lines are 1D model predictions. All simulations used 600 decelerator rings.
    }
\end{figure*}

Figure \ref{AllHex} shows the results of simulations and the 1D model for decelerating a buffer-gas beam from 180~m/s to 25~m/s for different hexapole lengths using a linear chirp. Here we vary the velocity of the index molecule, which determines the time the decelerator turns on and the magnitude of the acceleration. In these results, the amplitude of the 1D model for a linear chirp for a 1.5~m guide has been scaled to match the simulation peak height. The model predictions for all other curves were then scaled by this same factor.

    There are several important things to note. The first is the excellent agreement between the 1D model and the full 3D simulations. The 1D model reproduces the shape and relative amplitudes of the full simulations  results and allows us to explore a large parameter space in a short amount of time. For example, this 1D model is used to characterise non-linear chip deceleration that is very challenging to accurately test using our trajectory calculations. Second, the optimal index molecule velocity is not necessarily the mean speed of the pulse. In the case of a 1.5 m hexapole, the largest number of molecules are decelerated for $V_i = 210 m/s$. If 180 m/s is naively chosen, the resulting number of decelerated molecules is decreased by a factor of 2. Finally, one can see that for our beam parameters a 1.5 m hexapole produces the largest number of decelerated molecules, with $\sim 5\%$ stably decelerated to the target velocity.

\begin{figure}[ht]
    \includegraphics[width=\linewidth]{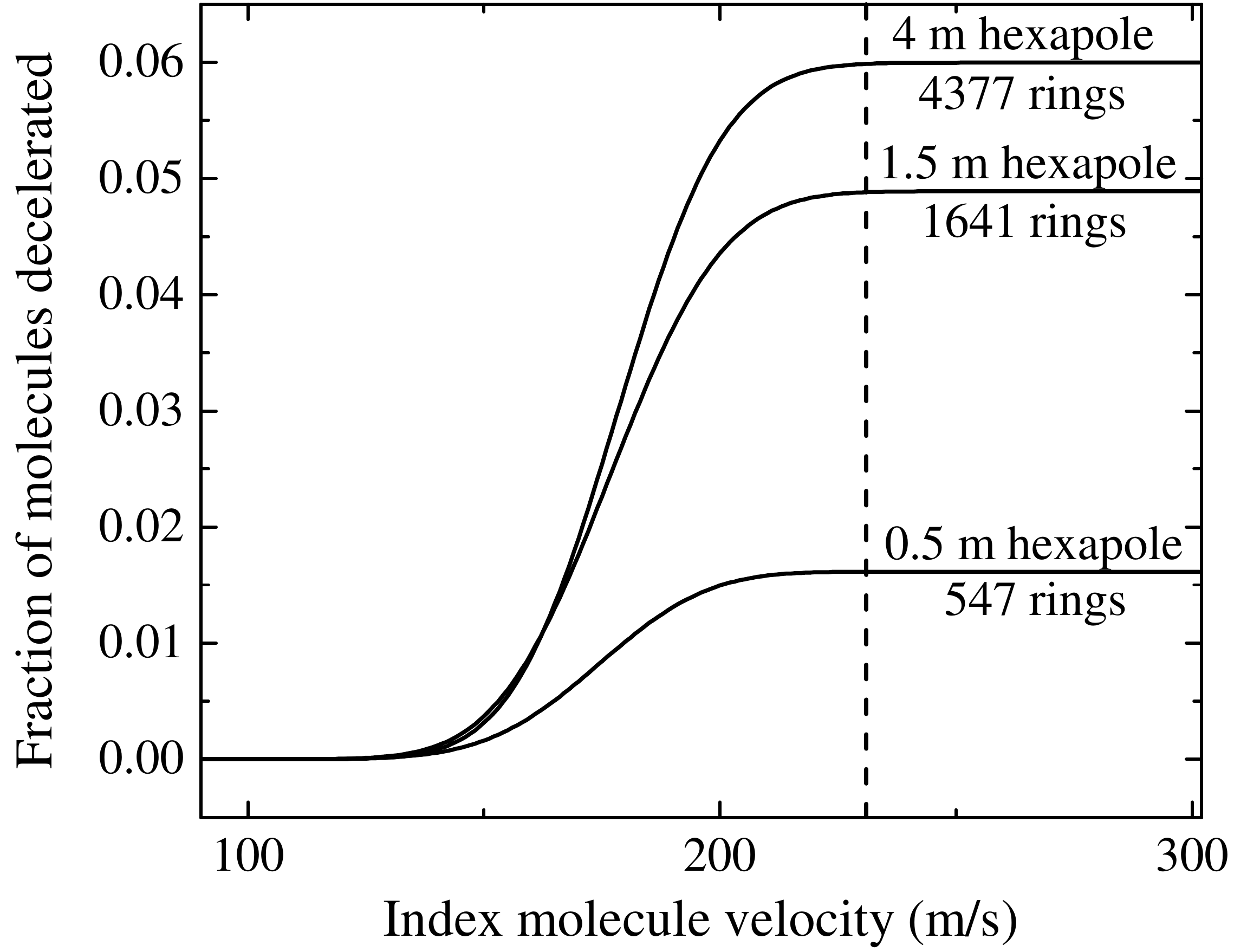}
    \caption{ \label{Modelcurves}
    1D model predictions of the fraction of molecules decelerated using the $1/t$ chirp for hexapole guide lengths of 0.5~m, 1.5~m, and 4~m. The number of rings required to decelerate the molecular pulse using a particular hexapole guide length are shown for a index molecule with a velocity of 231 m/s (dashed line), which is 180 m/s plus three times the longitudinal velocity width.
    }
\end{figure}

We also tested deceleration schemes using a deceleration function that matched the incoming velocity of the molecular beam (i.e., $H/t$) as discussed in Section III. The results of that model are shown in Fig. \ref{Modelcurves}. For the $1/t$ chirp scheme, the index molecule sets only the time the decelerator turns on. Thus, if it turns on before most of the molecules reach the decelerator entrance, the number of molecules decelerated should be independent of the exact index molecule velocity (decelerator turn on time).  This effect can be seen as a plateau in the number of molecules decelerated for large index molecule velocities (Fig. \ref{Modelcurves}). Comparing the number of molecules decelerated in the plateau region for various hexapole lengths, one can see the fraction increases with hexapole length. As the molecular beam becomes more correlated in position and velocity with longer hexapole lengths, the PSA will better match the incoming molecular beam PSD. The increase will saturate once the overlap of the PSD of the molecular beam and PSA of the decelerator is a maximum. This occurs at a fraction of 6\%. (Note: 90 \% of the molecular beam is outside the transverse PSA regardless of the correlation length.)

An important thing to note when evaluating deceleration schemes is the number of rings required for each scheme. Essentially any length may be chosen for a linear chirp scheme, with shorter decelerators requiring larger accelerations. However, for the $1/t$ chirp, the number of rings is a fixed value, which is a function of the hexapole length and the initial and final velocities. The number of rings required for a $1/t$ chirp is given by

\begin{equation}\label{chirpV}
N_{rings}= \frac{H\ln{\frac{V_i}{V_f}}}{\Delta d},
\end{equation}

where $\Delta d$ is the ring spacing. Thus, to realize the large gain in decelerated molecule fraction, one must build an unreasonably long Stark decelerator of several meters(Fig. \ref{Modelcurves}). If the length of decelerator is fixed at 600 rings, the linear chirp produces three times more decelerated molecules than the 1/t chirp.

    Until now, we have evaluated the deceleration schemes based on total number of molecules decelerated. For experiments that use a slow controlled molecular beam, the total number or integrated flux is the important metric, but for experiments requiring loading molecules into a trap, density also plays a role. A table of the decelerated fraction, molecular density, and number of rings used for the different protocols is shown Table II. The densities were calculated from the simulations by counting the number of molecules in the central well of the decelerator and assuming the molecules were uniformly distributed in the well volume.  We note this underestimates the true peak well density. We expect that slowing protocols that make use of longer hexapoles would result in decreased well densities because longitudinal phase-space distribution spreads during the flight time in the hexapole. This idea is borne out in the case of linear slowing protocols; the peak density decreased for longer hexapole length, although the decrease is not very significant over the range explored.

\begin{table*}[htb!]\label{DensityTable}
\begin{center}
\begin{tabular}{|c| c c c| c c |}
\hline
\multicolumn{1}{|c|}{Hexapole Length} &\multicolumn{3}{c|}{Linear Deceleration} &\multicolumn{2}{c|}{$\frac{1}{t}$ Deceleration}\\
& Density (mol./cc)&Fraction& Rings& Fraction&Rings\\
\hline
0.5m & $1.7\times10^9$ & 0.025&600& 0.016 & 547\\
1.5m &$1.5\times10^9$ & 0.045&600& 0.049 & 1641\\
4.0m &$8.2\times10^8$ & 0.025&600&  0.060 & 4377\\
\hline
\end{tabular}
\caption{The fraction of the initial beam that is decelerated to 25~m/s, central well densities, and number of decelerator rings used for various correlation (hexapole) lengths for both linear and $1/t$ acceptance functions. The number of rings used to decelerated using the $1/t$ chirp was set by the final velocity of 25~m/s. The density in the central well was calculated by assuming a uniform distribution within the well and thus represents a slight underestimate of the peak density.
}
\end{center}
\end{table*}

\section{Conclusions}
We have created an intense source ($2 \times 10^{11}$ /ablation pulse) of ground state CH radicals via buffer-gas cooling of a laser-ablated plume of iodoform. Using previously published cell extraction measurements, we estimated the parameters of a molecular beam that could be created from this buffer-gas source. We have shown that this extended pulse can be efficiently decelerated in a traveling-wave Stark decelerator by first correlating the position and velocity of molecules in the beam by guiding in an electrostatic hexapole. By optimizing this protocol using both 3D trajectory simulations and a 1D model we are able to decelerate 5\% of the molecules in the initial beam. The 1D model allows rapid exploration of a large parameter space of deceleration protocols and allows for understanding of the dynamics of matching the phase-space distribution of the beam with the phase-space acceptance of the decelerator. In the near future, we hope to be able to physically combine the buffer-gas source with the traveling-wave decelerator to explore combining the individual wells into a single high density and number electrostatic trap, which will allow for precise studies of collisions and reactions of CH radicals.

\begin{acknowledgments}
This material is based upon work supported by the National Science Foundation under Grant Numbers PHY 0900190, 1125844, 0748742, and 1265905, and the Air Force Office of Scientific Research under Grant Numbers FA9550-09-1-0588 and FA9550-12-1-0182. We gratefully acknowledge helpful discussions with John M. Doyle.
\end{acknowledgments}

\bibliography{CH_Paper}
\bibliographystyle{apsrev4-1}

\end{document}